\documentclass[aps,prl,twocolumn,superscriptaddress]{revtex4-2}

\usepackage{graphicx}
\usepackage{dcolumn}
\usepackage{bm}

\usepackage[utf8]{inputenc}
\usepackage[T1]{fontenc}
\usepackage{booktabs, array, mathptmx, float, tabularx, booktabs, lipsum, amsmath,multirow}
\usepackage{siunitx, xcolor}
\usepackage[version=4]{mhchem}
\usepackage{hyperref}

\begin{document}


\title{Observation of a $Pbca$ phase and robust metallicity in $\rm{RuO_2}$ under pressure}


\newcommand{\iopcas}{Beijing National Laboratory for Condensed Matter Physics, Institute of Physics, Chinese Academy of Sciences, Beijing 100190, China}
\newcommand{\ucas}{School of Physical Sciences, University of Chinese Academy of Sciences, Beijing 100049, China}
\newcommand{\yanshan}{Center for High Pressure Science (CHiPS), State Key Laboratory of Metastable Materials Science and Technology, Yanshan University, Qinhuangdao, Hebei, 066004, China}
\newcommand{\hpstar}{Center for High Pressure Science and Technology Advanced Research, 10 East Xibeiwang Road, Haidian, Beijing 100094, China}
\newcommand{\spring}{SPring-8/JASRI, 1-1-1 Kouto, Sayo-gun, Sayo-cho, Hyogo 679-5198, Japan}
\newcommand{\wollongong}{Institute for Superconducting and Electronic Materials, University of Wollongong, Innovation Campus, Squires Way, North Wollongong, NSW 2500, Australia}
\newcommand{\songshanlake}{Songshan Lake Materials Laboratory, Dongguan, Guangdong 523808, China}

\author{He Zhang}
\thanks{These authors have equal contributions.}
\affiliation{\iopcas}
\affiliation{\ucas}

\author{Xudong Wei}
\thanks{These authors have equal contributions.}
\affiliation{\yanshan}

\author{Wei Zhong}
\thanks{These authors have equal contributions.}
\affiliation{\hpstar}

\author{Xiaoli Ma}
\affiliation{\iopcas}

\author{Liyunxiao Wu}
\affiliation{\iopcas}
\affiliation{\ucas}

\author{Jie Zhou}
\affiliation{\hpstar}

\author{Saori Kawaguchi}
\affiliation{\spring}

\author{Hirokazu Kadobayashi}
\affiliation{\spring}

\author{Zhenxiang Cheng}
\affiliation{\wollongong}

\author{Guoying Gao}
\email[]{gaoguoying@ysu.edu.cn}
\affiliation{\yanshan}

\author{Xiaohui Yu}
\email[]{yuxh@iphy.ac.cn}
\affiliation{\iopcas}
\affiliation{\ucas}
\affiliation{\songshanlake}

\author{Ho-kwang Mao}
\affiliation{\hpstar}

\author{Binbin Yue}
\email[]{yuebb@hpstar.ac.cn}
\affiliation{\hpstar}

\author{Fang Hong}
\email[]{hongfang@iphy.ac.cn}
\affiliation{\iopcas}
\affiliation{\ucas}
\affiliation{\songshanlake}


\date{\today}

\begin{abstract}
    $\rm{RuO_2}$ stands as a quintessential rutile-type compound under ambient conditions, with its structural exploration under pressure bearing significant implications for both phase transition investigations and Earth science. Nonetheless, the precise phase transition sequence remains a debate. In this study, we disclose the emergence of the $Pbca$ phase alongside the enduring metallic character of $\rm{RuO_2}$ under megabar pressure. Employing state-of-the-art synchrotron X-ray diffraction, our observations delineate a phase transition trajectory progressing through rutile, $\rm{CaCl_2}$, and ultimately $Pbca$ phases. Notably, the $Pbca$ phase manifests immediately just after the rutile-$\rm{CaCl_2}$ transition, confining a narrow pressure regime for the pure $\rm{CaCl_2}$-type phase. Within the pressure range of 15.5 to 35.0 GPa, a coexistence of the $\rm{CaCl_2}$-type and $Pbca$ phases is observed, transforming to a sole presence of the $Pbca$ phase beyond 35.0 GPa. Electrical transport measurements conducted on both single crystal and powder samples confirm the enduring metallic conductivity of $\rm{RuO_2}$, persisting up to at least $\sim$120 GPa, albeit exhibiting a diminished conductivity at ultrahigh pressures due to a reduction in electronic density of states at the Fermi level. This study furnishes compelling evidence for the presence of the $Pbca$ phase across a broad pressure range, diverging from the previously widely acknowledged $Pa\bar{3}$ phase, thereby offering crucial insights into phase transition phenomena in other metal dioxides and advancing our comprehension of electronic behaviors within 4d and 5d electron systems.
\end{abstract}


\maketitle



Metal oxides exhibit a wide array of physical properties, encompassing ferroelectricity, magnetism, charge ordering, superconductivity, and topological physics \cite{RN640, RN641, RN642}. The majority of metal oxides tend towards insulating behavior, particularly alkali metal oxides, alkaline-earth metal oxides, and 3d metal oxides. However, an anomaly is observed in 4d and 5d metal oxides, as a significant portion of them display intrinsic metallicity \cite{RN643}. Among these metallic compounds, $\rm{RuO_2}$ stands out. Early investigations suggested $\rm{RuO_2}$ to adopt an undistorted rutile structure, with its Fermi surface topology being notably sensitive to spin-orbital coupling \cite{RN643}. However, neutron scattering studies have unveiled a distorted behavior in $\rm{RuO_2}$, with indications of room temperature antiferromagnetism \cite{RN644}. The origin and manifestation of this antiferromagnetism remain subject to debate; while previous studies advocate for a collinear antiferromagnetic arrangement \cite{RN645}, recent theoretical calculations posit a chiral signature in the magnetic symmetry of $\rm{RuO_2}$ \cite{RN646}. The observation of spontaneous or anomalous Hall Effect in $\rm{RuO_2}$ further complicates its structural and magnetic characteristics. This phenomenon is attributed either to the breaking of crystal time-reversal symmetry \cite{RN647} or to the altermagnetic character of $\rm{RuO_2}$ \cite{RN648}. Consequently, the understanding of the structure and magnetism of $\rm{RuO_2}$ proves to be more intricate than initially anticipated. Despite these complexities, theoretical propositions of a spin-splitter effect \cite{RN649} have been put forth, and experimental evidence of spin-splitter torque has been documented \cite{RN650, RN651}. Moreover, the generation of tilted spin currents from $\rm{RuO_2}$ thin films has been achieved \cite{RN652}, hinting at the potential of $\rm{RuO_2}$ as a novel platform for designing functional spintronic devices.

\begin{figure}[ht]
    \includegraphics[width =8.5cm]{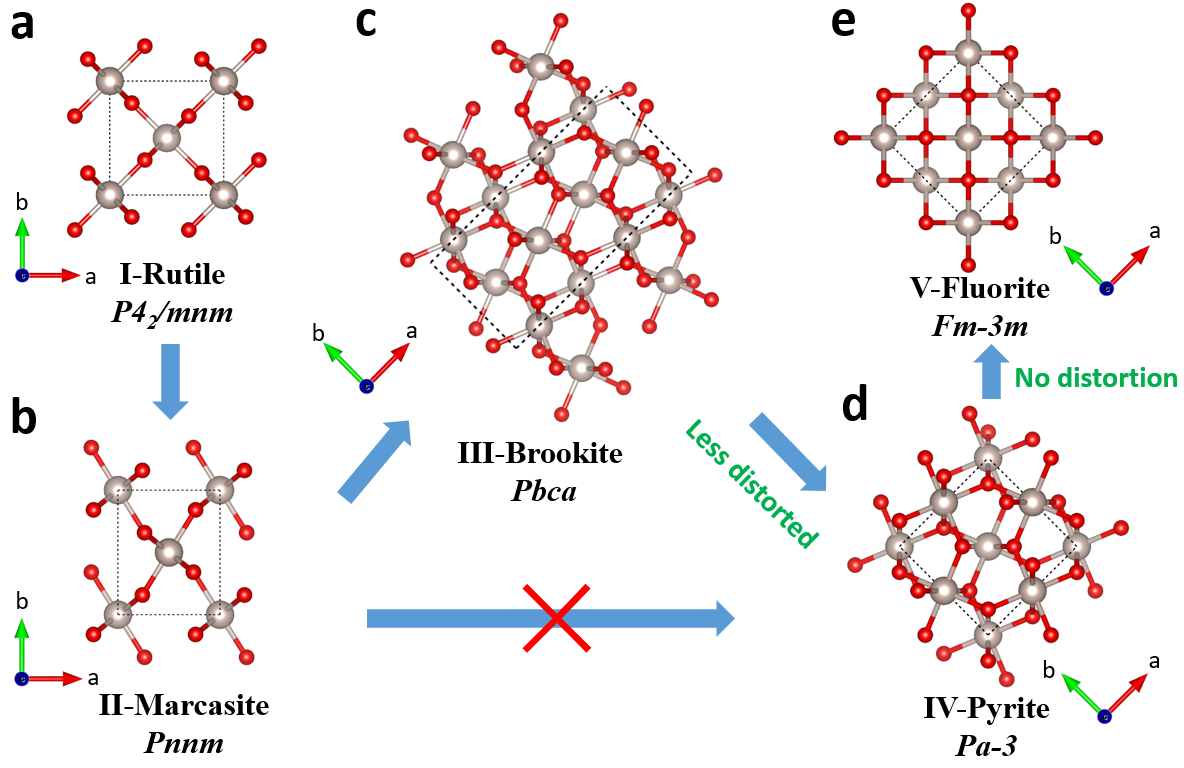}%
    \caption{\label{Fig1}The possible phase transition sequence of $\rm{RuO_2}$ under pressure. \textbf{a}, the original low-pressure Rutile phase (Phase I, $P4_2/mnm$); \textbf{b}, the intermediate Marcasite $\rm{CaCl_2}$-type phase (Phase II, $Pnnm$); \textbf{c}, the high-pressure Brookite phase (Phase III, $Pbca$); \textbf{d}, the high-pressure Pyrite phase (Phase IV, $Pa\bar{3}$); \textbf{e}, the high-pressure Fluorite phase (Phase V, $Fm\bar{3}m$). In the past several decades, the high-pressure phase is believed to be Phase IV in form of a cubic $Pa\bar{3}$ structure. In this work, the high-pressure phase is found to be a distorted $Pa\bar{3}$ structure, and the symmetry is lowered to the orthorhombic $Pbca$, which has a much larger lattice cell compared with that of $Pa\bar{3}$ phase.}
\end{figure}

In addition to its intriguing physical properties, the study of rutile $\rm{RuO_2}$'s structure and electronic behavior under pressure has attracted significant attention, owing to the prevalence of its typical rutile structure in $\rm{MO_2}$-type metal oxides such as $\rm{PbO_2}$, $\rm{SnO_2}$, and $\rm{MnO_2}$ \cite{RN653}. Given that many of these rutile-type oxides are natural minerals, research in this domain holds considerable importance for Earth science, and investigations into pressure-induced structural phase transitions in $\rm{RuO_2}$ are anticipated to shed light on analogous phenomena observed in $\rm{SiO_2}$ \cite{RN654}. Despite its seemingly straightforward chemical composition, the high-pressure structural and electronic characteristics of $\rm{RuO_2}$ remain contentious. In 1993, J. Haines et al. proposed a phase transition sequence for $\rm{RuO_2}$ comprising "Tetragonal rutile ($P4_2/mnm$)-orthorhombic marcasite/$\rm{CaCl_2}$ ($Pnnm$)-cubic fluorite ($Fm\bar{3}m$)" \cite{RN655}. Subsequent studies by the same group identified the high-pressure cubic phase as the $Pa\bar{3}$ phase through X-ray diffraction and neutron scattering, characterizing it as a distorted fluorite $Fm\bar{3}m$ phase \cite{RN656, RN657}. Consequently, a consensus emerged within the research community, with many scholars considering the high-pressure phase of $\rm{RuO_2}$ following the $\rm{CaCl_2}$-type phase to be the $Pa\bar{3}$ phase \cite{RN654, RN658, RN659, RN660}. This phase transition sequence appears plausible and parallels that of its analog compound $\rm{SnO_2}$ \cite{RN661}, albeit with the distinction that the $Pbca$ phase is stabilized after the $Pa\bar{3}$ phase in $\rm{SnO_2}$, with both phases regarded as distorted fluorite structures. Theoretical calculations have proposed the $Pbca$ phase for $\rm{RuO_2}$, albeit with indications of significant forces acting on the atoms that could compromise structural stability, particularly under high pressures \cite{RN658}. Recent experimental investigations on $\rm{RuO_2}$ have suggested the potential existence of the $Pbca$ phase, which, however, cannot be unequivocally distinguished from the $Pa\bar{3}$ phase, as both phases yield similar fitting results in structure refinement utilizing a multiple-phase model \cite{RN659}. Meanwhile, in the same work, they claimed that $\rm{RuO_2}$ becomes an insulator above 28 GPa. Although the insulating behavior is presumed to originate from the cubic fluorite phase, the absence of the fluorite phase in their XRD patterns under high pressure adds to the complexity of the situation \cite{RN659}. Consequently, a definitive consensus regarding the high-pressure phase of $\rm{RuO_2}$ remains elusive. 

In this study, we present compelling evidence confirming the presence of the $Pbca$ phase and the persistent metallic nature of $\rm{RuO_2}$ up to $\sim$120 GPa. The structural examination of $\rm{RuO_2}$ was conducted using state-of-the-art synchrotron X-ray diffraction techniques combined with theoretical structural searching methods, while the electronic behavior was investigated through electrical transport measurements and calculated electronic band structures. Our X-ray diffraction results uncovered an oversight in previous experimental studies, where an additional peak at a low diffraction angle had been disregarded or attributed to low-pressure phases or experimental artifacts  \cite{RN656, RN659}. Notably, structure refinement based on a multiple-phase model and discrete XRD patterns at limited pressure points may overlook crucial information pertaining to phase transitions \cite{RN656, RN659}. Our data delineate a transition sequence for $\rm{RuO_2}$, progressing through rutile, $\rm{CaCl_2}$ phase, and $Pbca$ phase, as depicted in Fig. \ref{Fig1}. Within a pressure range of approximately 15 to 35 GPa, a coexistence of the $\rm{CaCl_2}$-type and $Pbca$ phases is observed, with $\rm{RuO_2}$ transitioning to a pure $Pbca$ phase above 35 GPa. The $Pbca$ phase exhibits a significantly larger lattice cell compared to the $Pa\bar{3}$ or $Fm\bar{3}m$ phases, manifesting an additional diffraction peak at low angles. This peak weakens under higher pressures, indicating a potential transformation of $\rm{RuO_2}$ to a less distorted $Pa\bar{3}$ phase or the insulating $Fm\bar{3}m$ fluorite phase. Contrary to prior assertions regarding the conductivity of $\rm{RuO_2}$ under high pressure \cite{RN659}, our findings demonstrate its sustained metallicity over a wide pressure range, at least up to $\sim$120 GPa, as evidenced by measurements on single crystal $\rm{RuO_2}$. The previously observed insulating behavior can be attributed to strong electron scattering from grain boundaries in powder samples and pressure-induced distortion/defects under nonhydrostatic pressure conditions, resulting in a semimetal or semiconductor-like transport behavior. Transport and infrared reflectance measurements on powder samples further corroborate the metallic state of $\rm{RuO_2}$, at least up to $\sim$57 GPa, with minimal resistance changes indicating weak dependence on temperature and pressure. Theoretical calculations confirm the dynamic stability of the $Pbca$ phase, despite its higher enthalpy compared to the $Pa\bar{3}$ and $Fm\bar{3}m$ phases. Pressure-induced changes in the density of states near the Fermi level result in a reduced density of states above 80 GPa in the $Pbca$ phase, explaining the observed degeneration of conductivity under pressure. This study resolves the high-pressure structure of $\rm{RuO_2}$ and provides robust evidence for the presence of the $Pbca$ phase, challenging the previously accepted notion of the $Pa\bar{3}$ phase.

\begin{figure*}[t]
    \includegraphics[width=15cm]{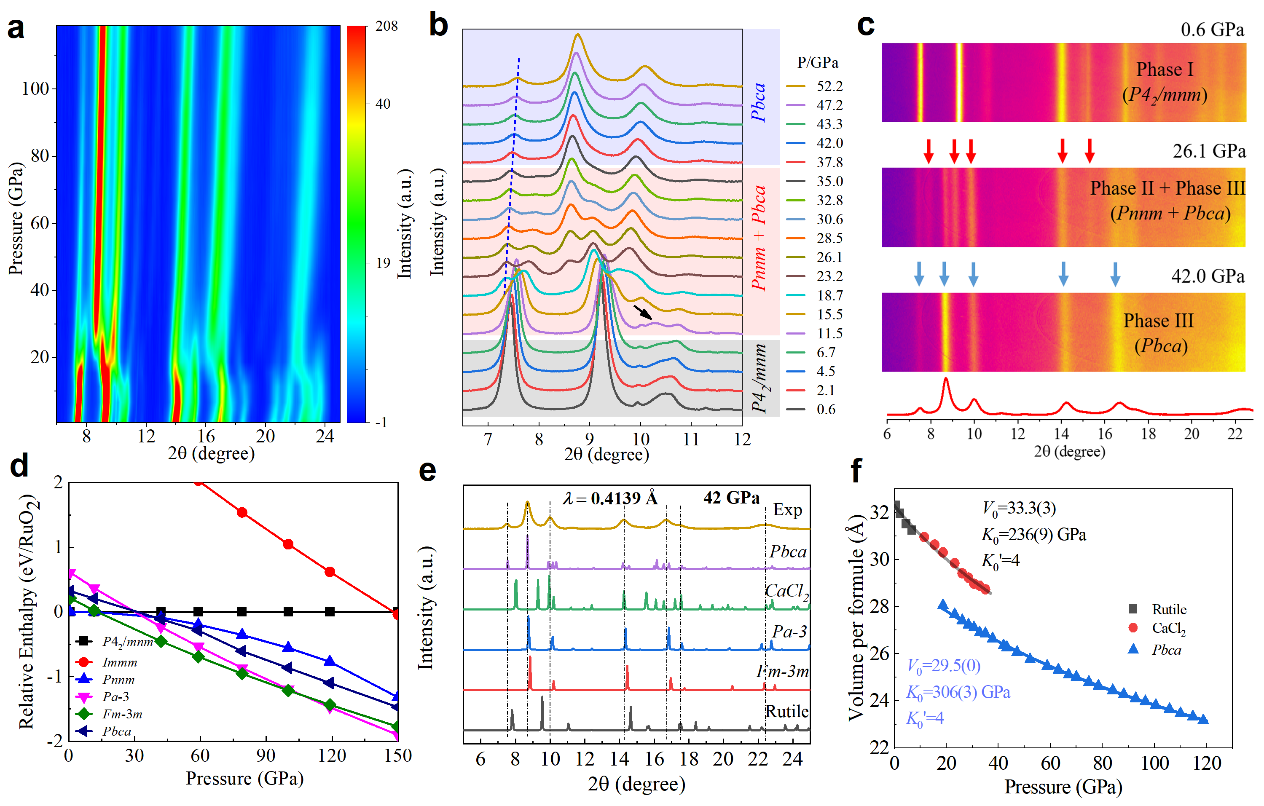}%
    \caption{\label{Fig2}The \textit{in situ} structure analysis of $\rm{RuO_2}$ by using synchrotron x-ray diffraction under pressure and calculated relative enthalpies for different structures. \textbf{a}, the contour plot of XRD patterns up to $\sim$120 GPa; \textbf{b}, the line plot of XRD up to 52.2 GPa with zoom-in region at low diffraction angles, and the phase boundaries are given; \textbf{c}, the original diffraction patterns for different phases, red arrows show the Bragg positions of Phase II ($Pnnm$) phase, and Phase II is absent at 42.0 GPa; \textbf{d}, the pressure dependent relative enthalpy for various phases; \textbf{e}, the comparison between experimental XRD and calculated XRD at 42 GPa, showing that $Pbca$ is the right structure rather than $Pa\bar{3}$ or $Fm\bar{3}m$; \textbf{f}, the equation of state of $\rm{RuO_2}$ under pressure up to $\sim$120 GPa.}
\end{figure*}

 \textit{In situ} high-pressure powder X-ray diffraction with a wavelength of 0.4199 \AA \  was carried out using synchrotron facilities at BL10XU beamline of SPring-8 to investigate the crystal structure of the samples, with pressure applied via a gas membrane system \cite{RN664}, and data reduction was completed by Dioptas software \cite{RN663}.Additionally, temperature-dependent electrical resistance measurements of both single crystal and powder $\rm{RuO_2}$ samples were conducted under high pressure and low temperature using the standard four-probe method. Pristine single crystal $\rm{RuO_2}$ was synthesized using a vapor transport technique \cite{RN662}.High-pressure infrared reflection spectroscopy was performed at room temperature to examine the spectral characteristics of the samples. More details can be found in Supplementary information. Structural relaxations and electronic properties were carried out using density functional theory with the Perdew-Burke-Ernzerhof parametrization of the generalized gradient approximation, as implemented in the Vienna ab initio simulation package \cite{RN665, RN666}. We adopted projector augmented wave method with $4s^24p^64d^75s^1$ for Ru and $2s^22p^6$ for O represented as the valence electrons \cite{RN667}.The cutoff energy is set to 800 eV and appropriate Monkhorst-Pack k meshes are used in the calculations. Phonon calculations were performed by the supercell method PHONOPY \cite{RN668, RN669}.

The structure of $\rm{RuO_2}$ under pressure was analyzed using synchrotron X-ray diffraction, employing a panel detector that offers enhanced spatial resolution compared to a point detector. The findings are illustrated in Fig. \ref{Fig2}a-c. Fig. \ref{Fig2}a presents a contour plot of all XRD patterns up to $\sim$120 GPa, while Fig. \ref{Fig2}b focuses on line patterns up to 52.2 GPa, with a zoom-in region highlighting low diffraction angles. A clear structural phase transition at 11.5 GPa, indicated by the arrow in Fig. \ref{Fig2}b, is identified as a second-order transition from rutile (Phase I) to marcasite $\rm{CaCl_2}$-type phase (Phase II), consistent with prior Raman studies (critical pressure: 11.8 GPa) \cite{RN670}. Another transition occurs above 15.5 GPa. This transition is characterized by an additional diffraction peak beyond the $\rm{CaCl_2}$-type phase, marked by the dash line in Fig. \ref{Fig2}b, persisting up to $\sim$120 GPa with a clear pressure-dependent shift to higher angles, suggesting intrinsic sample behavior rather than experimental artifacts. The high-pressure phase emerges immediately after the Phase I - Phase II transition near 11.5 GPa, leaving only a narrow pressure window for pure Phase II, consistent with previous experimental and theoretical findings \cite{RN655, RN656, RN658}. A wide coexistence region between 15.5 GPa and 35.0 GPa is observed, with the intensity of the high-pressure phase increasing with pressure.

Above 35 GPa, the XRD pattern does not show obvious change. Apart from the peak marked by the dash line, the high-pressure Phase III does look like the $Pa\bar{3}$ phase. However, such a peak was previously assigned to the Phase-II or experimental setup \cite{RN656, RN659}, due to the limited data. Here, it is clear that the high-pressure phase is in pure phase above 35 GPa, and it is not $Pa\bar{3}$ phase. For clarity, the original XRD patterns at representative pressures are displayed in Fig. \ref{Fig2}c, the Phase II ($Pnnm$) is completely absent at 42.0 GPa, excluding the possibility of $Pa\bar{3}$ coexisting with Pnnm. Considering the pattern similarity with $Pa\bar{3}$ phase, the high-pressure phase is expected to be closely related to $Pa\bar{3}$ and could be in a superstructure or subgroup of $Pa\bar{3}$ phase. To identify this phase, the $Pbca$ phase is relaxed and matches well with the experimental data, though the calculation shows a higher energy in $Pbca$ phase than $Pa\bar{3}$ and $Fm\bar{3}m$ phases above $\sim$30 GPa, as seen in Fig. \ref{Fig2}d-e. In spite of this, the energy of $Pbca$ phase is still below rutile phase above $\sim$30 GPa and $\rm{CaCl_2}$-type phase above $\sim$38 GPa. Actually, $Pbca$ phase and $Pa\bar{3}$ are both distorted $Fm\bar{3}m$ phase \cite{RN671}, as seen in Fig. \ref{Fig1}, and the difference between $Pbca$ and $Pa\bar{3}$ phase is their two nearest Ru-Ru bond distances: for $Pbca$ phase, the bond distances are slightly different while they are equal for $Pa\bar{3}$ phase. At 42 GPa, the lattice parameters of $Pbca$ phase are: $a$ = 9.5829 \AA, $b$ = 4.6741 \AA, $c$ = 4.7214 \AA. More structure information about various phases is provided in Table S1 in Supplementary information.The equation of state is displayed in Fig. \ref{Fig2}f based on the structural refinement results, and some representative XRD refinement patterns are displayed in Fig.S1 as shown in Supplementary information. For the Phase I-Phase II phase transition, there is no clear change of volume, confirmed the second order feature. On the contrary, there is a clear volume collapse when the high-pressure $Pbca$ phase is presented, showing the typical feature of a first-order structural phase transition. The structural correlation and evolution process among various phases can be well understood in Fig. \ref{Fig1}, Fig.S2 and Table S2 as shown in Supplementary information. For Phase I (rutile phase), the $\rm{RuO_6}$ hexahedron has 4 equal in-plane Ru-O bonds and 2 equal out-of-plane Ru-O bonds, the in-plan and out-of-plane Ru-O bonds are normal to each other but the bond distances are different. For Phase II ($\rm{CaCl_2}$-type phase), the nearest $\rm{RuO_6}$ hexahedrons rotate a little bit. For Phase-III ($Pbca$ phase), the sample becomes more dense and the distortion of $\rm{RuO_6}$ hexahedron is much stronger, in which the six Ru-O bonds are not equal to each other, as seen in Table S1. The less distorted $Pa\bar{3}$ or undistorted $Fm\bar{3}m$ phase may be favored and transform from $Pbca$ phase under higher pressure, as the typical peak of $Pbca$ phase (marked by dash line in Fig. \ref{Fig2}b) becomes weaker under ultrahigh pressure (as seen in Fig. \ref{Fig2}a).

\begin{figure}[ht]
    \includegraphics[width=8.5cm]{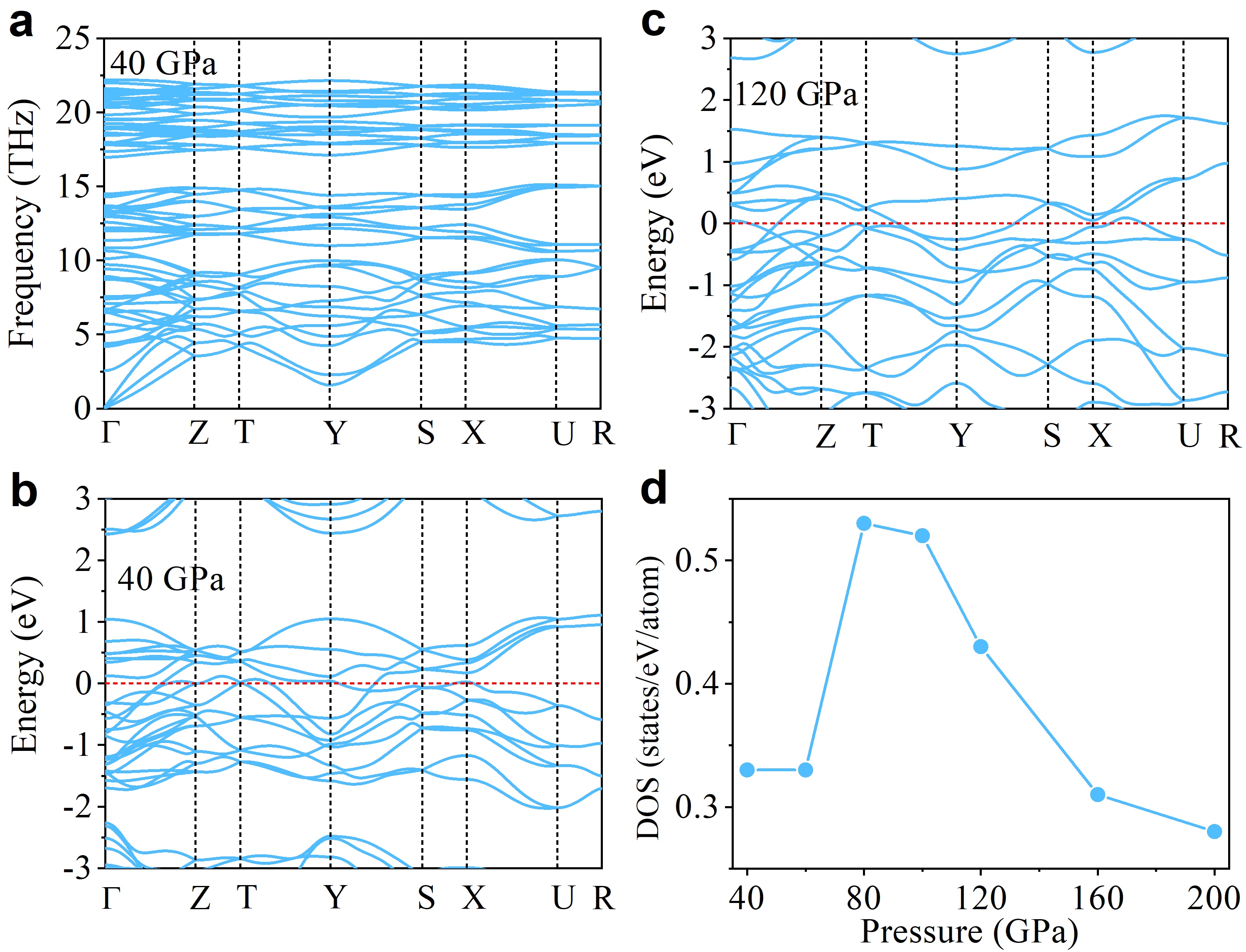}%
    \caption{\label{Fig3}Calculated phonon spectra and electronic behavior for the $Pbca$ phase under pressure. \textbf{a}, the phonon spectra at 40 GPa, and no imaginary frequency is presented, suggesting the dynamical stability of the $Pbca$ phase at this pressure; \textbf{b}, the electronic band structure of $Pbca$ phase at 40 GPa, showing a typical metallic feature; \textbf{c}, the electronic band structure at 120 GPa, still showing a metallic feature; \textbf{d}, the pressure dependent electronic density of state (DOS) at the Fermi level, showing a suppressed DOS under ultrahigh pressure.}
\end{figure}

The dynamical stability of the $Pbca$ phase is investigated by the phonon spectra, as presented in Fig. \ref{Fig3}a. There is no imaginary frequency in the whole phonon spectra at 40 GPa, indicating the dynamical stability of the $Pbca$ phase. It is also noted that earlier study claims the instability of $Pbca$ phase, but they cannot exclude the existence of this phase \cite{RN658}. To further explore the dynamical stability of the $Pbca$ phase under higher pressure, the phonon spectra at 200 GPa is also calculated (as seen in Fig.S3 in Supplementary information) and the results show that it is also dynamically stable at this pressure. Meanwhile, the electronic band structures of the $Pbca$ phase under pressure are calculated and the results show a metallic state over a large pressure range, as presented in Fig. \ref{Fig3}b-d. The calculated electronic DOS at the Fermi level shows a kink near 80 GPa, above which it declines with pressure, suggesting a reduced conductivity. The metallic behavior of the $Pbca$ phase is distinguished from the recently reported loss of metallicity above 28 GPa in a powder $\rm{RuO_2}$ sample\cite{RN659}. In the Ref. \cite{RN659}, they claimed a new high-pressure arsenopyrite-type phase with metallic behavior when $\rm{RuO_2}$ becomes an insulator. Such a statement is to some extent in conflict with each other, leaving an open question about the intrinsic electron behavior of $\rm{RuO_2}$ under high pressure. 

\begin{figure*}
    \includegraphics[width=14cm]{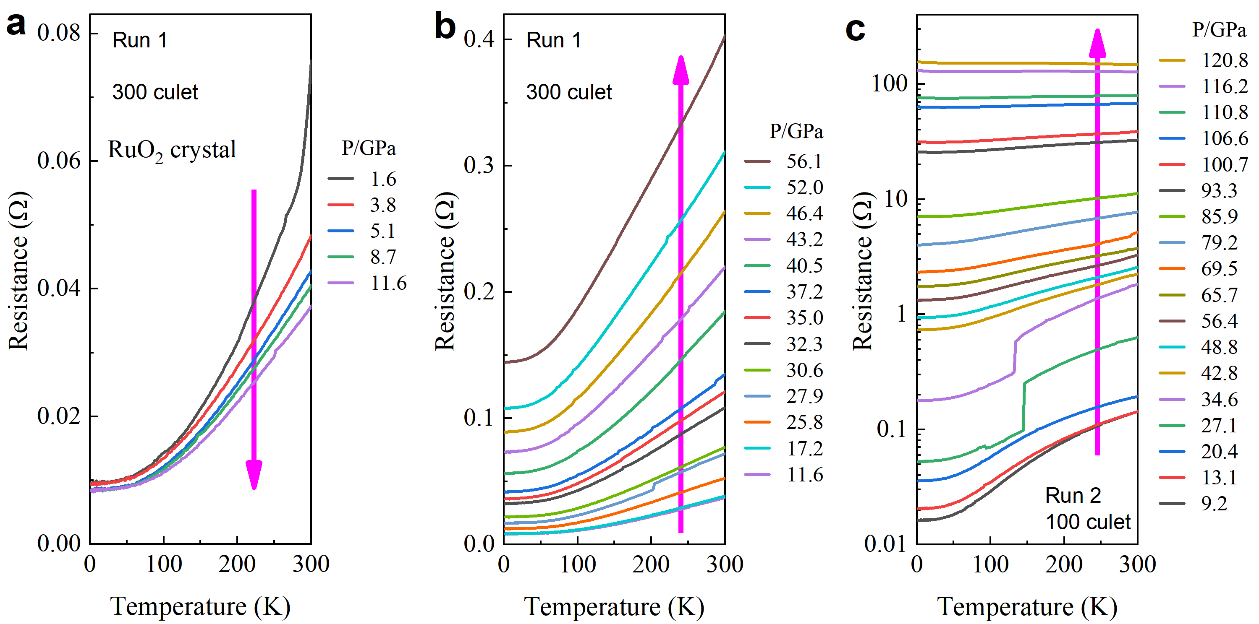}%
    \caption{\label{Fig4}The electrical transport properties of $\rm{RuO_2}$ single crystal under pressure up to $\sim$120 GPa. \textbf{a}, the R-T curves from 1.6 to 11.6 GPa, showing an enhanced metallic feature with pressure (Run 1, 300 micron culet); \textbf{b}, the R-T curves from 11.6 to 56.1 GPa, showing a degenerated metallic feature with pressure (Run 1); \textbf{c}, the R-T curves from 9.2 to 120.8 GPa by using a diamond anvil cell with 100 micron culets, further confirming the robust metallicity of $\rm{RuO_2}$ over a large pressure range, though the conductivity becomes worse with pressure.}
\end{figure*}

\begin{figure*}[ht]
    \includegraphics[width=14cm]{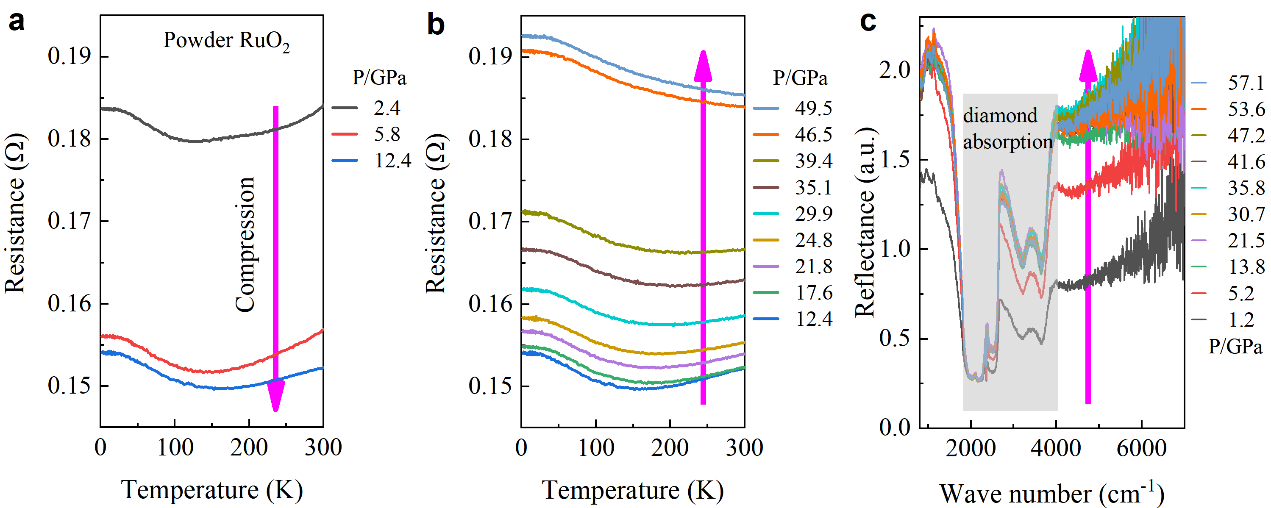}%
    \caption{\label{Fig5}The electrical transport and optical properties of $\rm{RuO_2}$ powder sample under pressure up to $\sim$57 GPa. \textbf{a}, the R-T curves from 2.4 to 12.4 GPa in powder $\rm{RuO_2}$, showing an enhanced metallic behavior with pressure, and the resistance shows weak temperature dependence; \textbf{b}, the R-T curves from 12.4 to 49.5 GPa in powder $\rm{RuO_2}$, showing a degenerated metallic feature with pressure and weak temperature dependence as well; \textbf{c}, the infrared reflectance of powder $\rm{RuO_2}$ from 1.2 to 57.1 GPa, confirming the robust metallic behavior in powder $\rm{RuO_2}$.}
\end{figure*}

To clear this confusion, the electrical transport measurement are done on both a single crystal sample and a powder sample. Fig. \ref{Fig4} displays the temperature dependent transport properties of a $\rm{RuO_2}$ single crystal under pressure up to $\sim$120 GPa. The measurements are carried out by two Runs. The Run 1 reaches a maximum pressure of 56.1 GPa by using a diamond anvil cell with 300-micron culets, while the Run 2 reaches 120.8 GPa by using 100-micron culets. In Run 1, the single crystal shows a slightly enhanced metallic feature with pressure at low pressure range up to 11.6 GPa, as seen in Fig. \ref{Fig4}a. The room temperature resistance changes a little bit more obvious than that at low temperature. Above 11.6 GPa, the resistance increases with pressure over the entire temperature range. Higher the pressure, larger the resistance. And such a trend is further confirmed by the results in Run 2 (Fig. \ref{Fig4}c). The pressure dependent resistance anomaly at 11.6 GPa is consistent with the second order structural phase transition from the rutile to the $\rm{CaCl_2}$-type phase. In Fig. \ref{Fig4}b, a clear resistance jump is presented at 27.9 GPa, and such a jump is also observed at 27.1 and 34.6 GPa in Run 2, as seen in Fig. \ref{Fig4}c. The pressure dependent resistance values at 300 K and 2 K also demonstrate such a resistance anomaly near 27.8 GPa, as shown in Fig. S4 in Supplementary information. The resistance jump is expected to stem from the Phase II-Phase III transition, as Phase III ($Pbca$) is less conductive than the Phase II due to a reduced DOS at the Fermi level (as shown in Fig.S5 in Supplementary information). Importantly, the resistance increases quickly with pressure, and the room temperature resistance at 120.8 GPa is more than three order of magnitude of that at 9.2 GPa, as seen in Fig. \ref{Fig4}c. In addition, at 120.8 GPa, the sample behaviors as a semimetal, since its resistance at low temperature is a little bit larger than that at high temperature. Such a trend is to some extent consistent with the calculation, as the DOS declines and $\rm{RuO_2}$ becomes less conductive (but is still in a metal state) above 80 GPa, though the experimental observation does not show such a critical pressure. Obviously, our transport data on single crystal provides an intrinsic picture about the robust metallicity of $\rm{RuO_2}$, rather than a loss of metallicity above 28 GPa in powder sample \cite{RN659}.

Whether the powder sample shows a different behavior from single crystal? To solve this puzzle, the electrical transport measurement on a powder sample is also conducted for reference, and the results is given in Fig. \ref{Fig5}. Similarly, the pressure dependent resistance shows an anomaly at 12.4 GPa, below which the resistance decreases with pressure, while it shows an increasing trend above 12.4 GPa. However, there is no resistance jump behavior during the Phase II - Phase III transition. Near ambient pressure, the big difference of R-T curves between powder and single crystal samples is their temperature dependence. For single crystal, a very standard metallic behavior is observed. For the powder sample, the R-T curve does follow a standard metallic behavior at high temperature, but it shows a semimetal behavior or “semiconducting” behavior at low temperature pressure, as seen in Fig. \ref{Fig5}a. Such an upturn trend at temperature range stems from the grain boundaries, which is often observed in polycrystalline samples and under high pressure with worse hydrostatic condition as it produces local distortions/defects and enhances the electron scattering \cite{RN672, RN673, RN674}. The semimetal behavior is more pronounced above 35 GPa but still shows weak temperature dependence, as seen in Fig. \ref{Fig5}b. It is noted that the resistance under various pressures does not show too much change, from $\sim$0.184 $\Omega$ at 2.4 GPa to $\sim$0.152 $\Omega$ at 12.4 GPa to $\sim$0.185 $\Omega$ at 49.5 GPa, suggesting a good metallic state. To exclude the effect of grain boundary scattering, optical method is used to further confirm the metallic behavior of powder $\rm{RuO_2}$, and the results are demonstrated in Fig. \ref{Fig5}c. At low pressure, such as 1.2 GPa, $\rm{RuO_2}$ is for sure in a metallic state and infrared reflectance shows a clear high value at low wavenumber range < 1800 $\rm{cm^{-1}}$. Upon compression, the reflectance increases and it almost does not change any more above 13.8 GPa. Therefore, the metallicity of powder $\rm{RuO_2}$ sample persists under pressure at least to 57.1 GPa, and our results exclude the possibility of an insulating transition in powder $\rm{RuO_2}$. The statement of insulating behavior in powder $\rm{RuO_2}$ in recent work should be due to the experimental setup, as no pressure medium is used rather than a hard insulating $\rm{Al_2O_3}$ layer \cite{RN659}, and the nonhydrostatic environment will amplify the effect of both the grain boundary and defect induced electron scattering in a large amount. In our work, we use KBr as the pressure medium and we don't see obvious color change as presented in Ref. \cite{RN659}. Actually, judging the conductivity or metallicity from the color is not reliable, though it works in most systems. 

Based on the electrical transport measurement on both a single crystal and a powder $\rm{RuO_2}$ sample, we establish with confidence that $\rm{RuO_2}$ maintains robust metallicity across a significant pressure range, extending at least to $\sim$120 GPa. However, the conductivity experiences a decline under ultrahigh pressures attributed to a reduction in the density of states at the Fermi level. Notably, meticulous structural analyses reveal that the high-pressure phase subsequent to the $\rm{CaCl_2}$-type phase is not the $Pa\bar{3}$ or $Fm\bar{3}m$ phases as previously presumed, but rather a $Pbca$ phase. This $Pbca$ phase, characterized by two unequal nearest Ru-Ru bonds, challenges conventional perceptions regarding the phase transition sequence in $\rm{RuO_2}$ cultivated over the past three decades. This discovery bears profound implications for both experimental investigations and theoretical calculations concerning rutile-type $\rm{MO_2}$ oxides. 

\begin{acknowledgments}
This work was supported by the National Key R\&D Program of China (Grants No. 2021YFA1400300), the Major Program of National Natural Science Foundation of China (Grants No. 22090041), the National Natural Science Foundation of China (Grants No. 12374050, No. 12004014, No. U1930401, No. 52372261). The \textit{in situ} XRD measurements were performed at BL10XU beamline (Proposal No. 2023B1412) at SPring-8 in Japan. Part of this work was partially carried out at the high-pressure synergetic measurement station of Synergic Extreme Condition User Facility.

F. H. and B.B.Y. conceived the project. G.G. and W.X. did the structural relaxations, phonon spectra and electronic band structure calculations. Z.X.C. provided the single crystals. W.Z., H. Z. and J.Z. collected the high pressure XRD. F.H., B.B.Y., X.L.M, and Y.W. did the analysis on the experimental structure data. X.H.Y. and H.Z. carried out the high pressure electrical transport measurement. H.K.M made comments on the results. F.H. wrote the manuscript with feedback from all other authors.

\end{acknowledgments}


\bibliography{reference}

\end{document}